\documentclass[aps,pre,reprint,showpacs,floatfix,superscriptaddress,longbibliography]{revtex4-1}

\usepackage{graphicx,bm,xcolor,amsmath}
\def \vec#1{{\mbox{\boldmath $#1$}}}

\begin{document}

\title{Finite volume solution for two-phase flow in a straight capillary}

\author{Alexander Yelkhovsky}
\noaffiliation
\author{W. Val Pinczewski}
\affiliation{School of Petroleum Engineering, University of New South Wales, 
Sydney 2052, Australia}

\begin{abstract}
The problem of two-phase flow in straight capillaries of polygonal cross section 
    displays many of the dynamic characteristics of rapid interfacial motions 
    associated with pore-scale displacements in porous media. 
Fluid inertia is known to be important in these displacements but is usually 
    ignored in network models commonly used to predict macroscopic flow properties. 
This study presents a numerical model for two-phase flow which describes the 
    spatial and temporal evolution of the interface between the fluids. 
The model is based on an averaged Navier-Stokes equation and is shown to be 
    successful in predicting the complex dynamics of both capillary rise in 
    round capillaries and imbibition along the corners of polygonal capillaries. 
The model can form the basis for more realistic network models which capture the 
    effect of capillary, viscous and inertial forces on pore-scale interfacial 
    dynamics and consequent macroscopic flow properties.
\end{abstract}

\maketitle

\section{Introduction}

The computational efficiency of pore network models has seen them increasingly 
    used to estimate important macroscopic multiphase flow properties for 
    porous media. 
The models are based on a simplified representation of the complex pore space 
    and simplified physics to describe the pore-scale configuration and motion 
    of the fluids in the pore space. 
Extensive reviews of the more recent developments in network modeling can be 
    found in Refs.~\cite{Blunt2013,Bondino2013,JoekarNiasar2012,Aghaei2015,
    Raeini2017}.

The more sophisticated network models represent the pore space as a 
    network of interconnected angular flow channels which retain some of the 
    more important geometrical and topological features of the actual pore 
    space and allow the presence of partially saturated pores where wetting 
    fluid is retained in corner films. 
The swelling of these films is important in imbibition where it results in 
    snap-off and the creation of residual nonwetting fluid 
    \cite{Nguyen2006,Idowu2010,Singh2017}.  
The movement of fluids in the network is assumed to occur in 
    a quasisteady sequence of equilibrium states governed by capillary 
    pressure (quasistatic models) or by a combination of capillary and viscous 
    forces (dynamic models). 
In dynamic network models the effect of viscosity is modeled as a perturbation 
    in capillary pressure determined from a pressure solution for steady-state 
    flow in the network.

A growing body of literature now exists questioning the validity of the 
    assumptions that the temporal evolution of fluid configurations can be 
    modeled by a sequence of equilibrium states and that the pore-scale fluid 
    motions leading to these states are steady. 
Recent studies of pore-scale displacements in two-dimensional micromodels using 
    high-speed photography \cite{Moebius2012,Armstrong2013,Moebius2014} 
    and in sandstone samples using fast x-ray computed micro-tomography 
    \cite{Berg2013,Berg2015,Rucker2015,Andrew2015,Armstrong2017} show that the 
    flow is inherently unsteady and that rapid interfacial motions associated 
    with pore-scale displacement events influence fluid topology and thus 
    macroscopic flow properties.
Well-known examples of those events are the wetting film snap-off induced by 
    the film swelling during imbibition, and the spontaneous displacement of 
    wetting fluid from a pore by nonwetting fluid from an adjoining pore throat 
    (Haines jump \cite{Haines}).

Moebius and Or \cite{Moebius2012,Moebius2014} show that in micromodel 
    displacements Haines jumps are associated with interfacial oscillations, 
    fluid rearrangements, and simultaneous cascade-like displacement events in 
    pores in the vicinity of the initial Haines jump. 
They attribute this behavior to the role of local inertial forces in pore-scale 
    displacements \cite{Moebius2014}. 
Armstrong and Berg \cite{Armstrong2013} report similar observations in their 
    micromodel experiments. 
They show that during the initial phase of a jump the interface is rapidly 
    accelerated to a peak velocity approximately three orders of magnitude 
    greater than the average displacement velocity. 
Moreover, they find no correlation between the interfacial velocities and the 
    mean displacement velocity. 
They also conclude that the displacements are dominated by fluid inertia at the 
    pore scale. 
Similar observations are reported by Berg {\it et~al.}~\cite{Berg2013} for 
    displacements in actual sandstones using high-speed x-ray computed 
    micro-tomography. 
They show that Haines jumps typically cascade through 10-20 pores and conclude 
    that the motion at the pore scale is capillary-inertial controlled with time 
    scales of the order of a millisecond consistent with acoustic measurements 
    reported by DiCarlo {\it et~al.}~\cite{DiCarlo2003}. 
R{\"u}cker {\it et~al.}~\cite{Rucker2015} describe similar rapid interfacial 
    motions for imbibition displacements in sandstone. 
They observed local snap-off events that caused meniscus oscillations and fluid 
    rearrangements in distant pores within the same cluster and concluded that 
    inertial forces are also important in imbibition.

The visualization studies are supported by computational studies on single pores 
    and small two-dimensional networks. 
Ferrari and Lunati \cite{Ferrari2014} used the Volume of Fluid method to 
    numerically solve the Navier-Stokes equations and track the evolution of the 
    interface in the corner of an angular pore. 
Their computations showed that local velocities can be orders of magnitude 
    greater than the injection velocity and that they induce damped oscillations 
    of the interface which depend only on fluid properties and pore geometry. 
The oscillations were of sufficient amplitude and duration to affect the order 
    of invasion and the macroscopic distribution of fluids in the network. 
Armstrong {\it et~al.}~\cite{Armstrong2015} used the density functional 
    hydrodynamics method to simulate their micromodel observations. 
The computed results were in agreement with their experimental observations, and 
    they conclude that temporal resolutions fine enough to resolve millisecond 
    events are required to fully capture the effects of the unsteady flow on 
    macroscopic flow properties.

The above studies clearly show that the flow of fluids in a network of 
    interconnected pores is inherently unsteady and that inertial effects during 
    rapid pore-scale displacements are important in determining the final 
    configuration of fluids in both drainage and imbibition.
Existing dynamic network models do not capture the unsteadiness of the flow and 
    neglect important inertial effects associated with rapid pore-scale fluid 
    reconfigurations. 
Meakin and Tartakovsky \cite{Meakin2009} suggest that these rapid fluid 
    reconfigurations also invalidate the capillary dominated quasistatic 
    assumption since dynamic (inertial) effects are also present in the limit of 
    slow flow.

The importance of inertial effects for the capillary rise phenomenon was previously 
    emphasized by Qu{\'e}r{\'e} {\it et~al.}~\cite{Quere1999}.
We agree with the authors of Ref.~\cite{Quere1999} that some experimental results
    cannot be explained without an account of inertia.
However, we disagree with the theoretical approach of Ref.~\cite{Quere1999}, which
    includes the energy conservation law applied to an open system,
    dissipation of flow energy without viscosity, and solution-dependent 
    equations of motion.

The present study describes a numerical model for the spatial and temporal 
    evolution of the interfaces in a capillary occupied by two immiscible fluids. 
The model can form the basis of a dynamic network model which includes 
    capillary, viscous, and inertial forces. 
This is demonstrated by applying the model to the problems of capillary rise in 
    a round tube whose end is brought into contact with the surface of a liquid, 
    and the imbibition along the corners of a polygonal capillary induced by 
    switching off the gravitational field.
Qu{\'e}r{\'e} \cite{Quere1997} reports experimental measurements of meniscus 
    position as a function of time in a round tube for three liquids of 
    different viscosities. 
The data display many of the characteristics of the dynamic pore-scale 
    displacements discussed above -- the importance of inertial forces 
    and the occurrence of damped oscillations if the liquid viscosity is low enough. 
For the second problem the model simulations are compared with data reported by 
    Weislogel and Lichter \cite{WeislogelLichter1998} for imbibition in 
    cells of polygonal cross section and the experiments reported by Dong and 
    Chatzis \cite{DongChatzis1995} for the flow of a wetting liquid along the 
    corners of a square capillary. 
For clarity the model and solution method are described for the case of a zero 
    contact angle. 
The general case of a dynamic contact angle is considered when the simulation 
    results are compared with the measured data.

\begin{figure}[ht]
    \includegraphics[width=40mm]{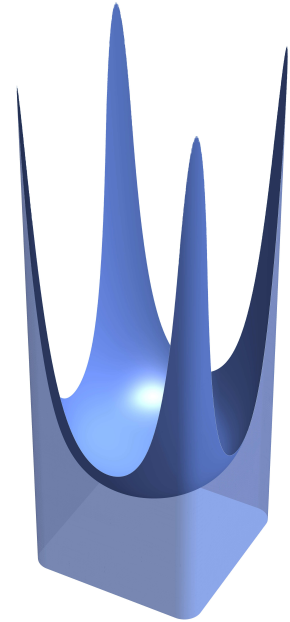}
    \caption{Sketch of the liquid phase configuration.}
    \label{interface}
\end{figure}

A sketch of the liquid phase configuration is shown at Fig.~\ref{interface} 
    for a tube with square cross section and rounded corners.
Below we discuss behavior of the liquid-gas interface in terms of its elements, 
    the meniscus (a rounded cap at the bottom of the interface), and the surface 
    of corner films protruding above the meniscus.
For a round tube there are no corner films, and the interface consists of only 
    the meniscus.

\section{Method}

To mimic a pore network, we represent the tube as a chain of fixed size flow 
    channels.
Assuming that the tube radius determines a characteristic scale of the flow, 
    we choose the  length of each channel to be of the order of the tube radius. 

Although this choice of discretization (as well as the choice of the variables 
    below) might be suboptimal for the particular problem of capillary flow
    in a tube of constant cross section, it is motivated by the prospect of 
    rather straightforward generalization of the method to proper networks.

In this section we outline our method according to the following sequence.
We start with a minimal set of variables describing the interface position
    at any moment in time.
Since the interface motion is governed by local fluxes, we derive the 
    corresponding equations of motion clearly stating the approximations made 
    on the way.
Our treatment of pressure differs from the approaches commonly used in network 
    models, and we devote a special subsection to explain the procedure.
We continue with descriptions of interface transitions between flow 
    channels and of the contact angle dynamics.
Finally, we conclude with brief notes on our numerical scheme.

\subsection{Interface}
 
The fluids are assumed to be incompressible. The evolution of the fluid-fluid 
    interface is governed by the volume balance equation,
\begin{equation}
    \label{volumeBalanceEq}
    \frac {dV}{dt} + \sum f = 0,
\end{equation}
    where $V$ is a subvolume of either of two phases, and $\sum f$ is total 
    outflux of the same phase through the subvolume boundaries.

We parametrize the interface geometry with a discrete set of variables 
    consisting of the following subsets.

Along the tube, bulk liquid is separated from bulk gas by the meniscus.
We define the meniscus position by an intersection of the spherical cap with 
    the tube inner surface.
Meniscus position along the tube comprises the first subset of interface 
    variables.
For a round tube, it is also the only one.

Behind the meniscus in an angular tube, we have gas in the bulk and liquid 
    film in the corners.    
If the film spreads over several flow channels, its curvature radii at the 
    channels joints comprise the second subset of interface variables.
Below we call these film hosting joints the anchors.

Finally, in a tube with rounded corners (like the one shown in 
    Fig.~\ref{interface}), a film can spread only up to the point where its 
    curvature reaches that of the corner, $1 / r_{\rm corner}$.
This point is called film tip, and its position comprises the third subset of 
    interface variables.

\begin{figure}[ht]
    \includegraphics[width=90mm]{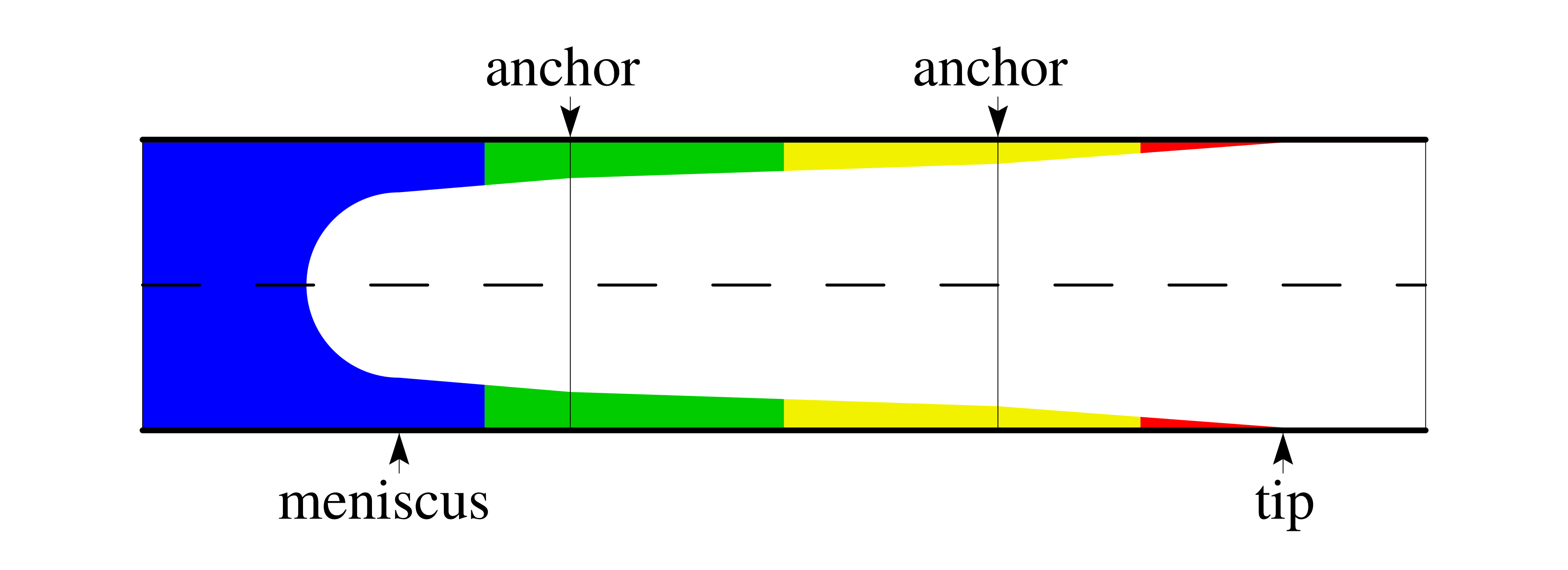}
    \caption{Longitudinal section of the flow. 
            Arrows point to discretization nodes. 
            Colors indicate the associated subvolumes.}
    \label{straight}
\end{figure}

Figure \ref{straight} shows part of the longitudinal section of the flow.
This section cuts along the axis (dashed line at the figure) and a corner of the 
    tube (shown as the thick black line).
The part of the tube shown in Fig.~\ref{straight} consists of three flow 
    channels of the same length; the left channel hosts the meniscus, the right 
    one hosts the tip, and the film in between is anchored by two joints.

To associate one balance equation (\ref{volumeBalanceEq}) with each variable, 
    we need one subvolume per a variable.
For an anchor, we draw cuts bounding its subvolume across the films midway 
    between the anchor and each of its closest neighbors (either another anchor 
    or the meniscus or the tip).
Then for an anchor or a tip, $V$ in Eq.~(\ref{volumeBalanceEq}) is the liquid 
    volume between the corresponding cuts, while $\sum f$ is a sum of liquid 
    outfluxes through those cuts.
On Fig.~\ref{straight}, the longitudinal section of the anchor subvolumes are 
    shown in green and yellow, while the meniscus and the tip subvolumes in blue 
    and red, respectively.

For the meniscus, $V$ is a volume of liquid between the meniscus and the cut 
    midway to the neighboring anchor, plus the part of the channel volume in 
    front of the meniscus.
We have to include the latter because it changes when the meniscus moves.
The total outflux for the meniscus is the liquid outflux through the cut plus 
    the liquid flux in front of and directed away from the meniscus.
By virtue of volume conservation, the latter equals the total flux and does not 
    depend on the cross section position.

When a flow starts, subvolumes associated with the interface variables start to 
    change, so that the meniscus and the tip start to move along the tube, while 
    the corner films start to swell.
At some moment, the meniscus or the tip can cross a boundary of a hosting 
    channel leading to a change in a number of anchors.
In a round tube, the meniscus just moves from one channel to another.

If subvolumes are chosen in the manner described above, we obtain a well-defined 
    system of coupled equations for the interface variables.
Volume balance is provided by the fact that for each cut an outflux for a 
    subvolume on one side is an influx for a subvolume on the other.

Once all subvolumes are known from the system of Eqs.~(\ref{volumeBalanceEq}), 
    we update values of all interface variables (meniscus and tip positions and 
    anchor radii) making up those subvolumes.
For a round tube, there is a trivial relation between the meniscus position and 
    a volume of liquid in front of it.
For an angular one, we also need to choose a smooth interpolation of the film 
    curvature between neighboring anchor points.
The linear behavior of capillary pressure is the most natural choice for this.

\subsection{Fluxes}

To close the system of balance equations (\ref{volumeBalanceEq}) we need to know 
    the liquid flux through each of the cuts as well as the total flux through 
    the tube. 

To calculate the fluxes, we start with the Navier-Stokes equation for an 
    incompressible fluid:
\begin{equation}
    \label{NavierStokesEq}
    \frac {\partial \vec v}{\partial t} + (\vec v \bm{\nabla}) \vec v 
    - \nu \Delta \vec v + \frac {\bm{\nabla} P}{\rho} - \vec g = \vec 0.
\end{equation}
Here $\vec v$ is the fluid velocity, 
    $P$ is the fluid pressure, 
    $\vec g$ is an apparent acceleration due to a body force (e.g., gravity),
    while $\rho$ and $\nu$ are the fluid density and kinematic viscosity, 
    respectively.
Here and below two vectors next to each other form the scalar product; the 
    Laplacian is defined in the standard way, $\Delta \equiv \bm{\nabla}^2$.

By definition, a fluid flux $f$ through a surface $\vec a$ is the integral 
    $\int {\rm d}^2 \vec a\, \vec v$ of fluid velocity $\vec v$ 
    relative to this surface.
The Einstein equivalence principle allows us to account for the surface motion 
    by substituting $\vec g \rightarrow \vec g - d \vec u / dt $, 
    where $\vec u$ is the surface velocity.

Integrating each term from the Navier-Stokes equation (\ref {NavierStokesEq}) 
    over a cut surface transverse to the channel axis, we get
\begin{equation}
    \label{masterEq}
    \dot f + 2\int {\rm d}a\, v v' - \nu \int {\rm d}a\, \Delta v 
    + \left( \frac {P'}{\rho} - g + \dot u \right) a = 0,
\end{equation}
    where $v$ is the fluid velocity component along the axis direction $\vec n$,
    $g = \vec g \vec n$ and $u = \vec u \vec n$,
    $a = \int {\rm d}a \equiv \int {\rm d}^2 \vec a\, \vec n$ is the surface 
    area, the dot and the prime denote partial derivatives over time and over 
    the axial coordinate, respectively.

If one phase fills a tube cross section completely, parameter $a$ in 
    Eq.~(\ref {masterEq}) is the area of this cross section, $a_{\rm tube}$.
It is convenient to introduce the dimensionless parameter $\Pi$ characterizing the 
    shape of the cross section, $\Pi = a_{\rm tube} / R^2$,
    where $R$ is the radius of the largest sphere fitting into the tube.

\begin{figure}[ht]
    \includegraphics[width=40mm]{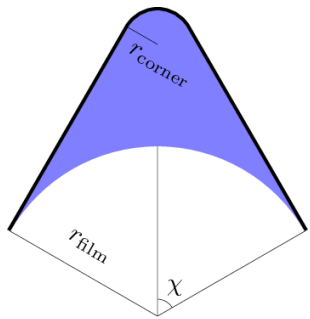}
    \caption{Geometry of the film flow cross section.}
    \label{corner}
\end{figure}

Then, if the liquid forms corner films, their cross sectional area 
    $a_{\rm film}$ is a function of the film curvature radius $r$.
For sharp corners, a dimensional consideration gives 
    $a_{\rm film} (r) = \Pi_{\rm film} r^2$, where $\Pi_{\rm film}$ is a 
    function of the corner angle.
For rounded corners, we subtract the area of the sharp tip 
    (see Fig.~\ref{corner}):
\begin{equation}
    a_{\rm film} (r) = \Pi_{\rm film} (r^2 - r_{\rm corner}^2),
\end{equation}
    so that $\Pi_{\rm film} = (\Pi - \pi) / (1 - r_{\rm corner}^2 / R^2)$ 
    according to the condition $a_{\rm tube} = \pi R^2 + a_{\rm film} (R)$. 

We would like to write an approximation for Eq.~(\ref {masterEq}) in terms 
    of flux and cross section geometry only.

First, the convective acceleration term $2\int {\rm d}a\, v v'$ can be safely
    neglected for the slow film flow.
For the bulk flow, we approximate it as $k\, f (f/a)'$ where $k$ is 2 for the 
    flat velocity profile and 8/3 for the hyperbolic (Poiseuille) one.

Then, for the velocity $v$ we use the standard approximation of the steady-state 
    flow profile, i.e., a solution of the 2D Poisson equation 
    $\Delta_{\perp} v = {\rm const}$,
    where $\Delta_{\perp}$ is the Laplacian in two dimensions perpendicular 
    to the flow.
The transverse viscosity term $-\nu \int {\rm d}a\, \Delta_{\perp} v$ can then 
    be rewritten as $\nu \beta f / s^2$, where $s$ is a characteristic size of 
    the flow cross section (radius of the tube for the bulk, film curvature 
    radius for the corner flow), and the flow resistance factor $\beta$ is a 
    dimensionless function of the cross section geometry encoding information 
    about the steady-state solution \cite{RansohoffRadke1988}.
This factor can be calculated as $a / \int {\rm d} a\, w$, where $w$ is the 
    solution to the equation $\Delta_{\perp} w = -1 / s^2$ with nonslip boundary 
    condition at the wall and the full slip condition at the gas-liquid interface.

\begin{figure}[ht]
    \includegraphics[width=86mm]{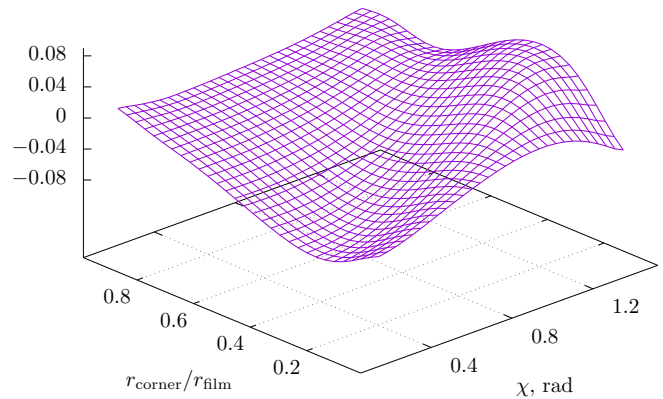}
    \caption{$1 - \beta_{\rm appr} / \beta_{\rm num}$ 
            for the film flow resistance function $\beta$.}
    \label{beta}
\end{figure}

For two particular cross section geometries of angular tubes discussed in this 
    paper, we could interpolate $\beta$ between values calculated at several points
    of $\zeta = r_{\rm corner} / r_{\rm film}$.
In a general case of a network with channels of various geometries, it is more 
    efficient to have an analytical approximation for $\beta$ as a function of both
    $\zeta$ and the cross section angle $\chi$:
\begin{equation}
    \label{betaApp}
    \beta_{\rm appr} = \frac {4.4}{c} 
    \left( \frac {1}{h_x^2} + \frac {1}{h_y^2} \right), 
\end{equation}
    where $c = 1 - 0.37 (1 - \zeta)^2 (1 + 0.2 \sin (2\chi))$, 
    $h_y = (\sec (\chi) - 1) (1 - \zeta)$, 
    $h_x = \tan (\chi/2)$ if $2 \zeta \cos^2 (\chi/2) < 1$ and 
    $\sqrt{h_y (2\zeta - h_y)}$ otherwise.
Figure \ref{beta} shows the relative difference 
    $1 - \beta_{\rm appr} / \beta_{\rm num}$ between the approximate solution 
    (\ref{betaApp}) and its numerical counterpart calculated with the help of 
    the PLTMG package \cite{pltmg}.

In contrast to the corner flow resistance factor which increases indefinitely 
    for thin films, its counterparts for the bulk and single-phase flows both 
    vary in finite ranges of the order of the round tube value $\beta = 8$.

The last integral entering Eq.~(\ref{masterEq}) we need to deal with is the 
    longitudinal part of the viscous term, $- \nu \int {\rm d}a\, v''$.
It vanishes for the single phase flow intervals since the fluids are 
    incompressible and the channel cross section does not change along its axis.
For the two-phase intervals, we use the incompressibility equation,
    $v' + \bm{\nabla}_{\perp} \vec v_{\perp} = 0$,
    together with the Stokes theorem to rewrite this term as 
    $\pm \nu \int {\rm d}b\, \dot r' = \pm 2 \nu \Pi_{\rm film} r \dot r'$,
    where $r$ is the film curvature radius at the cross section. The integral 
    is taken along the cross section boundary and the sign in front is plus 
    for the corner flow and minus for the bulk flow.

Summing up, the approximate form of the Eq.~(\ref{masterEq}) is
\begin{equation}
    \label{apprEq}
    \dot f + k f \left( \frac {f}{a} \right)' 
    + \nu \beta \frac {f}{s^2} \pm 2 \nu \Pi_{\rm film} r \dot r' 
    + \left( \frac {P'}{\rho} - g + \dot u \right) a = 0.
\end{equation}

\subsection{Pressure}

To close the system of Eqs.~(\ref{apprEq}), we need to calculate pressure 
    distributions along the flow.
It is sufficient to find the pressure on the tube axis; pressure at the corners 
    is determined by the condition $P_{\rm liq} = P_{\rm gas} - P_{\rm c}$, 
    where $P_{\rm c}$ is a local capillary pressure of the film.
Note that due to variation of the capillary pressure along the film, pressure 
    gradients $P'_{\rm liq}$ and $P'_{\rm gas}$ can have opposite signs inducing 
    counterflow of the two fluids.

Bulk pressure (that on the tube axis) can be calculated as follows.
First, for a cross section located at a point $z$ along the tube we can apply 
    Eq.~(\ref{apprEq}) to the total flux $F$.
For a single-phase cross section, $F$ is the flux of a phase at this cross 
    section, for a two-phase one, $F = f_{\rm liq} + f_{\rm gas}$.
Separating the total flux time derivative and bulk pressure gradient terms from the 
    rest and explicitly indicating $z$ dependence, we get:
\begin{equation}
    \label{xRateEq}
    \dot F = -c (z) P'_{\rm bulk} (z) + \delta (z).
\end{equation}
If a section crosses a film, both $c$ and $\delta$ are the sums of bulk and corner 
    contributions, otherwise they contain bulk contribution only (either of the 
    liquid or of the gas).

Volume conservation requires that the total flux $F$ is the same for 
    all cross sections of the tube.
Hence, dividing Eq.~(\ref {xRateEq}) by $c (z)$ and integrating along the tube 
    axis from the entrance to the exit, we get
\begin{equation}
    \label{totalRateEq}
    \dot F \int_{\rm entr}^{\rm exit} {\rm d}z \frac{1}{ c (z) } 
    = P_{\rm bulk}^{\rm entr} - P_{\rm bulk}^{\rm exit} 
    + \int_{\rm entr}^{\rm exit} {\rm d}z \frac{ \delta (z) }{ c (z) },
\end{equation}
    where $P_{\rm bulk}^{\rm entr}$ and $P_{\rm bulk}^{\rm exit}$ are the 
    bulk pressure values at the tube entrance and at the exit, respectively.
The pressure difference above is the sum of three terms: liquid pressure drop 
    between the entrance and the meniscus, gas pressure drop between the 
    meniscus and the tube exit, and the jump between the two which is equal to 
    the meniscus capillary pressure.

For a tube touching the surface of a liquid bath we also need to take into 
    account a flow of the liquid in the bath \cite{Szekeley}.
Assuming that this flow has only a radial component which does not depend on a 
    direction inside the bath, we can integrate the radial component of the 
    Navier-Stokes equation (\ref{NavierStokesEq}) along the radial coordinate 
    to obtain
\begin{equation}
    \label{bathEq}
    \dot F + \frac {1}{2 a R} F^2 - \frac {a}{\rho R} p - \frac {2 \nu}{3 R^2} F 
    = 0,
\end{equation}
    where $\rho$ and $\nu$ are density and kinematic viscosity of the liquid, 
    $R$ and $a$ are radius and area of a smallest hemisphere still allowing the 
    above approximations for the flow,
    and $p$ is a pressure difference between infinity and this hemisphere.
We assume that $R$ can be taken equal to the tube radius, while for $a$ we use 
    the value $\pi R^2$ (instead of $2\pi R^2$) and neglect a difference between 
    $p$ and a pressure at the tube entrance.

Finally, to simulate flows at a constant applied pressure, we use 
    Eqs.~(\ref{totalRateEq}) and (\ref{bathEq}) with the constraint 
    $p - P_{\rm bulk}^{\rm entr} + P_{\rm bulk}^{\rm exit} = 0$ to find the 
    total flux rate throughout the evolution.
To simulate flows at a constant flux rate, we use Eq.~(\ref{totalRateEq}) 
    to calculate the pressure drop between the tube ends. 
The pressure distribution along the tube can then be found from Eq.~(\ref{xRateEq}).

\subsection{Topology}

When the meniscus or the tip reaches a channel end, it disappears from that 
    channel and reappears in the next one.
If no new anchor appears as a result of this event, volume conservation is 
    sufficient to find a new position of the meniscus or the tip.
Otherwise, to determine a curvature radius at the new anchor we use the 
    condition that the capillary pressure gradient is the same on both sides of 
    the anchor.

\subsection{Contact angle}

Application of the above model to the capillary rise at the conditions of the 
    experiments \cite{Quere1997,DongChatzis1995,WeislogelLichter1998}
    quickly makes it apparent that an agreement with some of the measurements 
    can only be achieved if we consider nonzero contact angles.
Moreover, for the round tubes we also need to include the contact angle dynamics.

According to molecular kinetic theory, the correlation between contact angle 
    $\theta_{\rm d}$, its value at rest $\theta_{\rm s}$, and the contact line 
    velocity $w$ has the form (see Ref.~\cite{Blake2013} and references therein)
\begin{equation}
    \label{dynamicAngle}
    \cos \theta_{\rm d} - \cos \theta_{\rm s} + \xi\, {\rm Ca} = 0.
\end{equation}
Here ${\rm Ca} = w \nu / \sigma$ is the contact line capillary number, 
    $\nu$ is the liquid kinematic viscosity, 
    and $\sigma$ is the liquid-gas interface tension.
The contact line friction factor $\xi$ cannot yet be calculated at the present 
    state of the theory, and we use it as a fitting parameter to match our 
    simulation results with experimental data.

An alternative correlation, based on the hydrodynamic analysis of the moving 
    contact line problem \cite{Voinov1976,Cox1986},
\begin{equation}
    \label{VoinovCox}
    \theta_{\rm d}^3 - \theta_{\rm s}^3 - \kappa \, {\rm Ca} = 0,
\end{equation}
    where $\kappa$ is another friction factor to be used as a fitting parameter,
    gives slightly worse match of our results against experimental data,
    most probably because some of the flows have Reynolds number 
    Re $\sim 10^2$, 
    while the Voinov-Cox theory \cite{Voinov1976,Cox1986} is applicable at 
    Re $< 1$ only.

\subsection{Numerical scheme}

We compute the evolution of the interface using the fully implicit scheme with an  
    adaptive time step.
Each liquid subvolume is updated according to the balance equation 
    (\ref{volumeBalanceEq}).
After extracting new values of the interface variables (including a new contact 
    angle if necessary) we calculate a new pressure distribution as explained 
    above.
The liquid fluxes through the cuts between subvolumes are updated according to 
    Eq.~(\ref{apprEq}) (as  well as the total flux if necessary).
The iterations continue until new values of all variables are stable within 
    tolerance limits.

To calculate the integrals from Eq.~(\ref{totalRateEq}) we divide each flow 
    channel into single- and two-phase intervals.
For the flows considered in the present paper, each channel contains either two 
    intervals if it hosts the meniscus or the tip, or just one interval 
    otherwise.
Then we approximate the integrals from Eq.~(\ref{totalRateEq}) using the 
    rectangle rule applied to those intervals, for example,
\begin{equation}
    \label{approximateResistanceEq}
    \int_{\rm entr}^{\rm exit} {\rm d}z \frac {1}{c (z)} \approx \sum_{i} 
    \frac{ l_{i} }
    { a_{{\rm bulk},i} / \rho_{\rm bulk} + a_{{\rm film},i} / \rho_{\rm film} },
\end{equation}
    where $l_{i}$ is interval length, 
    while $a_{{\rm bulk},i} = a_{\rm tube} - a_{{\rm film},i}$ 
    and $a_{{\rm film},i}$ are the areas of the flow cross sections through the 
    middle of the interval, for the bulk and the film, respectively.

The simulations described below were carried out on a laptop with a Core 2 Duo 
    processor and 8 Gb of memory.
Most of the simulations take just a few second to run.
The only exception is the Soltrol 100 imbibition process (see below), which 
    took a few hours both to observe \cite{DongChatzis1995} and to simulate.

\section{Results}

\paragraph{Round capillary}

\begin{figure}[htb]
    \includegraphics[width=86mm]{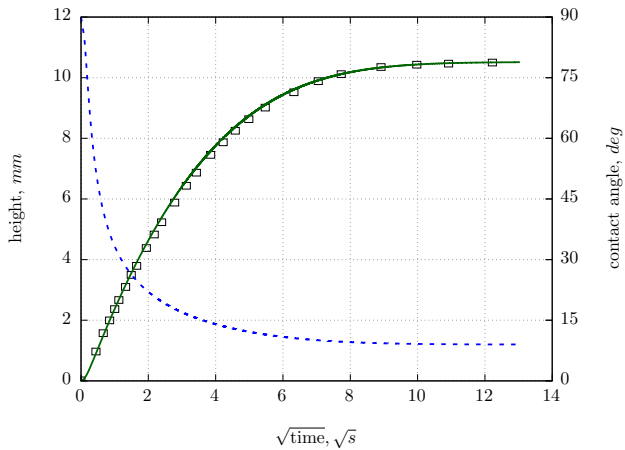}
    \caption{Capillary rise of silicone oil 
        (density 980 kg/m$^3$, viscosity 500 cP, 
        air interface surface tension 21.1 mN/m) 
        in a glass tube of radius 421 $\mu$m.
        Rectangles are experimental data for the rise height 
        from Ref.~\cite{Quere1997}, and solid and dashed lines are 
        simulation results for the height and dynamic 
        contact angle, respectively.}
    \label{silicone}
\end{figure}

\begin{figure}[htb]
    \includegraphics[width=86mm]{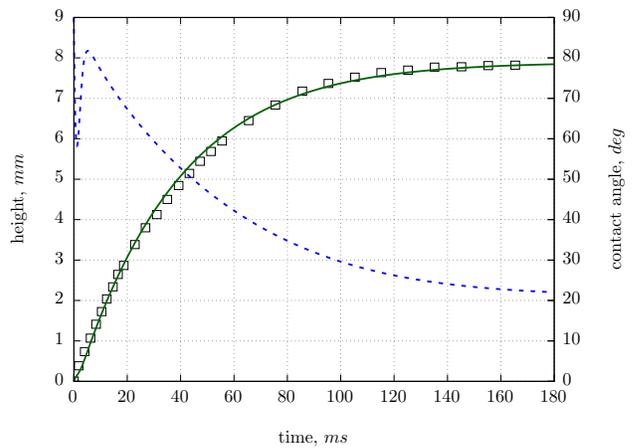}
    \caption{Capillary rise of ethanol 
        (density 780 kg/m$^3$, viscosity 1.17 cP, 
        air interface surface tension 21.6 mN/m) 
        in a glass tube of radius 689 $\mu$m.
        Notations are the same as in Fig.~\ref{silicone}.}
    \label{ethanol}
\end{figure}

\begin{figure}[htb]
    \includegraphics[width=86mm]{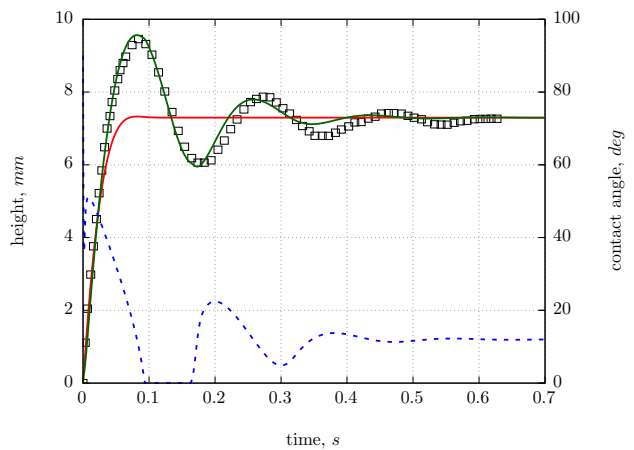}
    \caption{Capillary rise of ether 
        (density 710 kg/m$^3$, viscosity 0.3 cP, 
        air interface surface tension 16.6 mN/m) 
        in a glass tube of radius 689 $\mu$m.
        Notations are the same as in Fig.~\ref{silicone}.
        Red line is calculated with the inertial term suppressed by 1/5.}
    \label{aether}
\end{figure}

We use measured heights of capillary rise for three liquids (viscous silicone 
    oil, ethanol, and ether) from Ref.~\cite{Quere1997}.
Comparisons between our computations and the data are shown in 
    Figs.~\ref{silicone}-\ref{aether}.

To match our results with experimental data we use the following procedure.
First, the static contact angle $\theta_{\rm s}$ is extracted from the 
    observed equilibrium height.
Then, we run a few simulations with this $\theta_{\rm s}$ and various
    values of $\xi$ to get the initial slope of the capillary rise curve
    to match the observed one.
We do not try to achieve the best possible match; our aim is to demonstrate
    that the model captures major physical features of the process.

It is not evident from Ref.~\cite{Quere1997} how the meniscus height is defined.
The simulation results are presented for the height of the contact line above 
    the surface of the liquid bath.
We obtain the best agreement between our results and the measured data if we 
    assume nonzero static contact angles $\theta_{\rm s}$, approximately 
    9$^{\circ}$ for the silicone oil, 22$^{\circ}$ for ethanol, 
    and 12$^{\circ}$ for ether.
Also, we find that the best agreement with the measurements is achieved when 
    the friction factor $\xi$ from (\ref{dynamicAngle}) is approximately 5 
    for the silicone oil, 55 for ethanol, and 60 for ether.

Due to its low viscosity, the ether column overshoots its equilibrium height 
    and then recedes.
According to Ref.~\cite{Bretherton1961}, behind the meniscus a receding column 
    trails a film with thickness 3.72$\,{\rm Ca}^{2/3}R$, where $R$ is the tube 
    radius.
This film effectively increases the meniscus curvature.

To demonstrate importance of inertia, we ran a simulation with the flux time
    derivative term multiplied by 1/5 
    (the convection term is zero because velocity profile is the same for all 
    $z$'s).
The result shown in red at Fig.~\ref{aether} clearly indicates that inertia
    is responsible for the height oscillations.

\paragraph{Polygonal capillary}

In a series of experiments described in Ref.~\cite{DongChatzis1995}, a 4 to 
    5 cm long slug of a wetting liquid was introduced from one end of the 
    half-meter long tube with the square cross section.
Then both ends of the tube were sealed, the tube was turned from vertical to 
    horizontal position, and the motion of both the liquid-gas meniscus and the 
    film tip was recorded.

We simulate the flow of Soltrol 100 through the tube with the transverse size of 
    500 $\mu$m and rounded corners with radius of 51.2 $\mu$m.
The best agreement with the experimental data is achieved with the contact angle 
    $\theta_{\rm s} = 2.5^{\circ}$.
\begin{figure}[htb]
    \includegraphics[width=86mm]{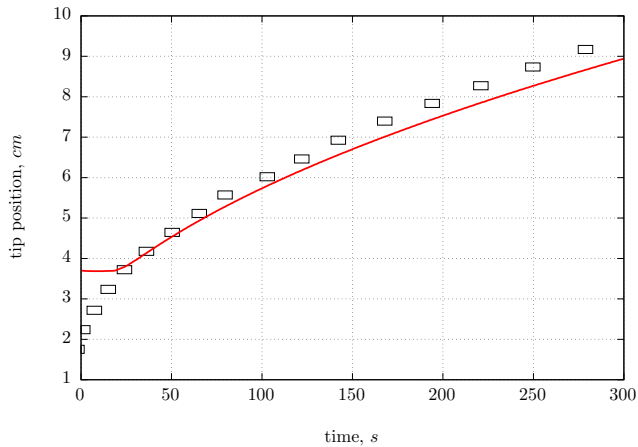}
    \caption{Imbibition of Soltrol 100 
        (density 739 kg/m$^3$, viscosity 0.98 cP, 
        air interface surface tension 22.0 mN/m) 
        along the corners of the 500 $\mu$m square glass tube.
        Rectangles are experimental data for the tip displacement 
        from Ref.~\cite{DongChatzis1995};
        solid line is our simulation result.}
    \label{soltrolEdge}
\end{figure}
Figure \ref{soltrolEdge} shows a comparison between the simulations and the 
    measured data for the displacement of the film tip.
We see a reasonable agreement after approximately 30~s from the start of the flow.
Looking at the film profile for the earlier times, we see a swelling wave 
    propagating from the meniscus and finally reaching the tip.
Initial distance from the meniscus to the tip is denoted by $x_0$ at 
    Ref.~\cite{DongChatzis1995}. 
This distance can be calculated from the equation
    $(\rho_{\rm liq} - \rho_{\rm gas})\, g\, x_0 = 
    \sigma / r_{\rm corner} - \sigma / r_{\rm film}$,
    where $r_{\rm film}$ is the film radius at the meniscus.
For Soltrol 100 we get $x_0 \approx$ 3.7 cm.
It is not clear from Ref.~\cite{DongChatzis1995} how exactly the tip position
    was measured to produce $x_0 \approx$ 1.7 cm.
Moreover, Fig.~6 from Ref.~\cite{DongChatzis1995} also shows a rather 
    different value of $x_0\approx$ 1.2 cm for Soltrol 170 even though 
    the relevant properties of two liquids are quite close: 0.739 versus 
    0.792 g/cm$^3$ for densities, and 22 versus 24 dyne/cm for surface 
    tensions, respectively.

\begin{figure}[htb]
    \includegraphics[width=86mm]{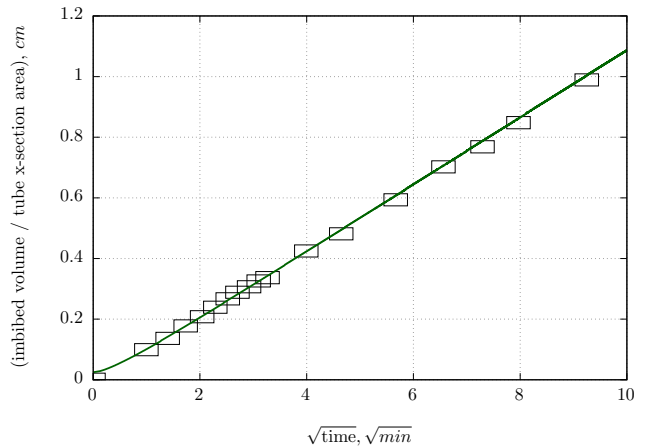}
    \caption{Same flow as at Fig.~\ref{soltrolEdge}, 
        measurement data and simulation result for the absolute 
        value of the meniscus apex displacement.}
    \label{soltrolFront}
\end{figure}
Figure \ref{soltrolFront} shows a comparison of our simulation result with the 
    data for meniscus displacement.

In the experiments \cite{WeislogelLichter1998}, acrylic test cells of polygonal 
    cross section were partially filled with a silicone oil and then dropped 
    from a 27 m high tower.
Interface evolution during the 2.2 second free fall was filmed at 128 frames 
    per second and then digitized.

We simulate the conditions for the ST10 experiment from 
    Ref.~\cite{WeislogelLichter1998} with 10 cS (9.35 cP) silicone oil inside 
    the 12 mm equilateral triangle tube.
\begin{figure}[htb]
    \includegraphics[width=86mm]{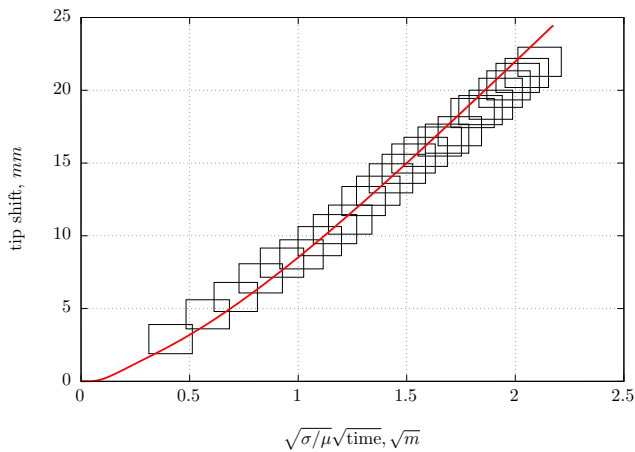}
    \caption{Imbibition of silicone oil 
        (density 935 kg/m$^3$, viscosity 9.35 cP, 
        air interface surface tension 20.1 mN/m) 
        along the corners of triangle acrylic tube with 12 mm side.
        Rectangles are experimental data for the tip displacement 
        from Ref.~\cite{WeislogelLichter1998};
        solid line is our simulation result.}
    \label{wledge}
\end{figure}
\begin{figure}[htb]
    \includegraphics[width=86mm]{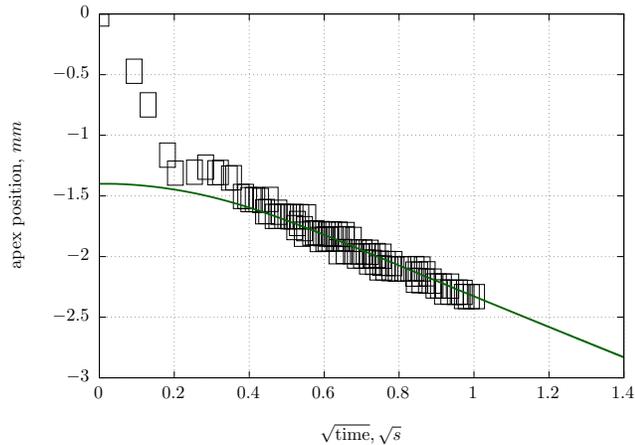}
    \caption{Same flow as at Fig.~\ref{wledge}, measurement data and 
        simulation result for the meniscus apex position.}
    \label{wlapex}
\end{figure}
A comparison between the simulated and measured data is shown in 
    Figs.~\ref{wledge} and \ref{wlapex}.
The agreement between the two looks quite reasonable except for the meniscus 
    apex position during the first approximately 0.16 seconds of zero-gravity 
    conditions.
According to Ref.~\cite{WeislogelLichter1998}, behavior of the system 
    at the earlier times is governed by relaxation of the meniscus shape from
    almost flat initially (due to the gravity field) to that of a spherical
    cap (due to the surface tension).
Since our primary interest is in small capillaries where gravity can be 
    neglected, we do not account for the effect of gravity on meniscus shape.

Figure 12 from Ref.~\cite{WeislogelLichter1998} (reproduced at Fig.~\ref{WL12}) 
    shows similar behavior observed in their experiment LT2.
According to Ref.~\cite{WeislogelLichter1998}, the film profile behavior in 
    general and an appearance of the constant height point in particular,
    are universal properties of the film flow along the tube corners.

\begin{figure}[t]
    \includegraphics[width=86mm]{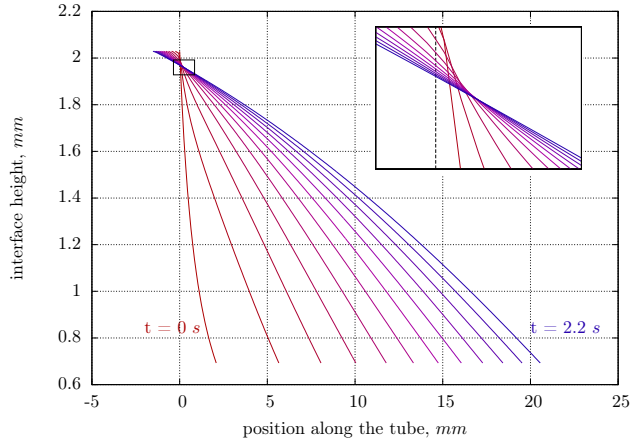}
    \caption{Simulated profile of the interface height along the tube.
        Snapshots are made every 0.2 seconds.
        The insert shows a close-up of the constant height point.}
    \label{profile}
\end{figure}
Figure \ref{profile} shows how the film profile changes with time. 
In particular, after approximately 0.6 seconds from the start of the flow, 
    all profiles come through the same point.
The height of this point, 1.96 mm, agrees well with the experimental value 
    1.97$\pm$0.08 mm \cite{WeislogelLichter1998} (see the insert in 
    Fig.~\ref{profile}).
\begin{figure}[htb]
    \includegraphics[width=86mm]{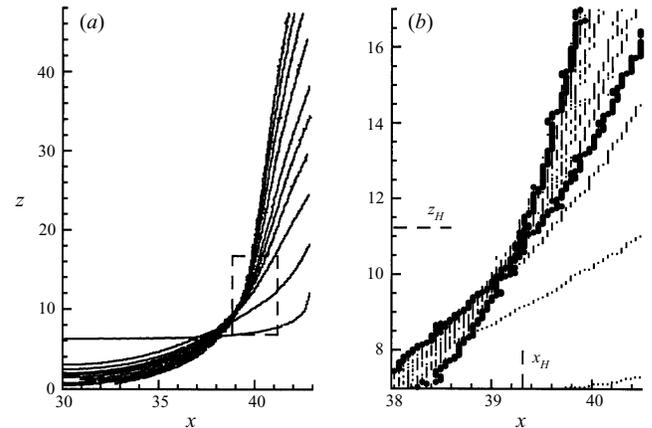}
    \caption{Interface profiles measured in one of the experiments from 
        Ref.~\cite{WeislogelLichter1998}.
        Interface height is measured along $x$, and $z$ is position 
        along the tube.}
    \label{WL12}
\end{figure}

\section{Conclusions}

A simple one-dimensional numerical model, based on an averaged Navier-Stokes 
    equation, is successful in predicting the dynamics of the flow in capillaries 
    of circular and angular cross section.
The inclusion of meniscus position as a model parameter results in an efficient 
    and accurate front-tracking scheme with sufficient temporal resolution to 
    resolve the initial acceleration of the interface and subsequent 
    inertia-driven oscillations.     
The mechanistic similarity between the onset of pore-scale interfacial 
    displacements in porous media and capillary flow
    \cite{Moebius2012,Moebius2014} suggests that the numerical model can form 
    the basis for more realistic network models which capture the effects of 
    capillary, viscous and inertial forces on pore-scale interfacial dynamics 
    and consequent macroscopic flow properties.

%

\end{document}